\documentclass[referee]{raa}            

\usepackage{graphicx,times}             
\usepackage{natbib}
\usepackage{amssymb,amsmath}
\bibpunct{(}{)}{;}{a}{}{,}

\usepackage[a4paper=true,dvipdfm=true,pagebackref=true]{hyperref}
\hypersetup{colorlinks = true, linkcolor = green, anchorcolor = red, citecolor = blue, filecolor = red, pagecolor = red, urlcolor = red}

\begin{document}

   \title{Detection of a high-confidence quasi-periodic oscillation in radio light curve of the high redshift FSRQ PKS J0805-0111
}

   \volnopage{Vol.0 (20xx) No.0, 000--000}      
   \setcounter{page}{1}          

   \author{Guo-Wei Ren
   	\inst{1,2}
   	\and Hao-Jing Zhang
   	\inst{1}
   	\and Xiong Zhang
   	\inst{1}
	\and Nan Ding
	\inst{3}
    \and Xing Yang
   	\inst{1}
    \and Fu-Ting Li
   	\inst{1}
    \and Pei-Lin Yan
   	\inst{1}
    \and Xiao-Lin Xu
   	\inst{1}
   }

   \institute{College of Physics and Electronics, Yunnan Normal University, Kunming 650500, P.~R.~China; {\it kmzhanghj@163.com}\\
\and
 Department of Astronomy, Xiamen University, Xiamen 361005, P.~R.~China;\\
        \and
             School of Physical Science and Technology, Kunming University, Kunming 650214, P.~R.~China;\\
\vs\no
   {\small Received~~20xx month day; accepted~~20xx~~month day}}

\abstract{ In this work, we have searched quasi-periodic oscillations (QPOs) in the 15 GHz light curve of the FSRQ PKS J0805-0111 monitored by the Owens Valley Radio Observatory (OVRO) 40 m telescope during the period from January 9,2008 to May 9,2019, using the weighted wavelet Z-transform (WWZ) and the Lomb-Scargle Periodogram (LSP) techniques. This is the first time to search for periodic radio signal in the FSRQ PKS J0805-0111 by these two methods. All two methods consistently reveal a repeating signal with a periodicity of 3.38$\pm$ 0.8 years ($>$99.7\% confidence level). In order to determining the significance of the periods, the false alarm probability method was used, and a large number of Monte Carlo simulations were performed. As possible explanations, we discuss a number of scenarios including the thermal instability of thin disks scenario, the spiral jet scenario and the binary supermassive black hole scenario, we expected that the binary black hole scenario, where the QPO is caused by the precession of the binary black holes, is the most likely explanation. FSRQ PKS J0805-0111 thus could be a good binary black hole candidate. In the binary black hole scenario, the distance between the primary black hole and the secondary black hole is about $1.71\times 10^{16}$ cm.
}

   \authorrunning{Guo-Wei Ren et al. }            
   \titlerunning{High-confidence radio QPO in FSRQ PKS J0805-0111}  

   \maketitle

%
%
\section{Introduction}           
\label{sect:intro}

Active galactic nuclei (AGNs) are very energetic extragalactic sources and they are generally powered by accreting supermassive black holes (SMBHs) with masses of $10^{6}-10^{10} M_{\odot}$ in the centers of the galaxies \citep{2019MNRAS.484.5785G, 2015A&A...576A.122E}. Blazars represent an extreme subclass of radio-loud AGNs with their relativistic jets aligned very closely to observers' line of sight, and they are characterized by large amplitude, rapid and violent variability across the entire electromagnetic spectrum, high and variable polarization at radio and optical energies, with non-thermal continuum emission ranging from radio to high-energy ¦Ã-rays, and superluminal jet speeds \citep{1995PASP..107..803U, 1980ARA&A..18..321A, 2017ApJS..229...21X}. Blazars are often subclassified into two categories according to the observed features: BL Lacertae objects (BL Lac) and flat spectrum radio quasars (FSRQs). BL Lac have featureless optical spectra with weak or no emission lines, possibly due to the emission being dominated by the jet, while FSRQ have a flat radio spectrum with a spectral index $\alpha \leqslant 0.5$ and broad quasar-like emission lines in the optical spectra. The blazars' emission is dominated by relativistic jets, and the beaming effect boost the relativistic jets \citep{2016AJ....151...54S}. The blazars' broadband spectral energy distributions (SEDs) have an obviously double-peaked structure. These two peaks have different physical origins: The low-energy peak at the IR-optical-UV band is considered to be caused by the synchrotron emission of relativistic electrons, and the high-energy peak at the GeV-TeV gamma-ray band is explained by the inverse Compton (IC) scattering \citep{2017ApJS..229...21X, 2007Ap&SS.309...95B, 1995ApJ...446L..63D}.

Blazars' light curves generally display a series of features, especially aperiodic or quasi-periodic variability in a wide range of temporal frequencies-equivalently. Blazars show signatures of QPOs in the multi-frequency blazar light curves, including radio, optical, X-ray and $\gamma$-ray, have been found \citep{2018Galax...6..136B}. The QPOs on different timescales, from decades down to a few minutes \citep{2018Galax...6....1G, 2017ApJ...847....7B, 2016ApJ...832...47B}. Research on QPO of blazars is one of the most active fields of extragalactic astronomy, and provides an important way to explore the radiation mechanism in blazars \citep{2018Ap&SS.363..169L}. According to the time spans of the variability, the characteristic timescales of variabilities can be broadly divided into three classes, viz., intraday variability (IDV) or micro-variability, which is defined as having timescales ranging from minutes to a few hours, short-term timescale variability (STV), which has timescales of days to weeks, even months, and long-term timescale variability (LTV) of a few months to years \citep{2016MNRAS.458.1127G, 2017ApJS..229...21X, 2018Ap&SS.363..169L}. The discovery of QPO in the light curve variability could have deep consequences on the global understanding of the sources, constituting a fundamental building block of models \citep{2016AJ....151...54S}.  It is need more long-term observations to search for periodic variations in many timescales.

The blazar PKS J0805-0111 (RA=08h 05m 12.9s, Dec= -01d 11m 37s; and Z = 1.388) was identified as a FSRQ, this source has been detected by many currently available instruments, it is cataloged by Fermi/LAT as 3FGL J0805.2-0112 \citep{2015ApJS..218...23A}, its optical (R-band) brightness was recorded to be 17.87 magnitudes \citep{2008ApJS..175...97H}. Since 2008, PKS J0805-0111 has been monitored in the 15GHz radio band by the 40 m telescope of the OVRO \citep{2011ApJS..194...29R}.

In this paper, we analyze the long term (~11.3 years) observations of FSRQ PKS J0805-0111 and report our discovery of a high confidence QPO of the 15 GHz radio flux variability. In section 2, we describe the 15 GHz radio observations. In section 3,
We present time series analysis of the light curve using the WWZ and the LSP, we test these two methods with astronomical analogue signals, besides, we also elaborate on the false alarm probability and the Monte Carlo simulation technique which is used to compute the statistical significance of the detected periodicity. In section 4, we discuss three scenarios to explain the QPO behavior of FSRQ PKS J0805-0111, and estimate the distance between the primary black hole and the secondary black hole; Our conclusions be summarized in section 5.


\section{Observations}
\label{sect:Obs}

OVRO announced 15GHz radio band observation data from a source FSRQ PKS J0805-0111 observed with a 40-meter telescope from January 9, 2008, to May 9, 2019, for a total of 4138 days ($\sim$11.3 years), a total of 507 data points. We conducted a preliminary analysis of these data. In the 15GHz radio band data of this source, the minimum flux is 0.14 Jy, the maximum flux is 0.76 Jy, the average flux is 0.44 Jy, and the standard deviation is 0.14. In order to quantify the observed variability, we estimate the variability amplitude (VA), which represents the peak-to-peak oscillation, and the fractional variability (FV), which represents the mean variability. Estimate the amplitude of the peak-to-peak variation using the relationship given in \cite{1996A&A...305...42H}:

\begin{equation}
VA=\sqrt{\left ( A_{max}-A_{min} \right )^{2}-2\sigma ^{2}}
\end{equation}

Where $A_{max}$, $A_{min}$ and $\sigma$ represent the maximum value, minimum value, and average value of the magnitude errors in the radio observation data, respectively. Similarly, we can calculate the fractional variance FV with a mean flux of $\left \langle F \right \rangle$ with $S^{2}$ variance, and $\left \langle \sigma _{err}^{2} \right \rangle$ is the mean error squared, as stated in the formula given in \cite{2003MNRAS.345.1271V}:

\begin{equation}
F_{var}=\sqrt{\frac{S^{2}-\left \langle \sigma _{err}^{2} \right \rangle}{\left \langle F \right \rangle^{2}}}
\end{equation}

The error of fractional variability can be estimated as stated in the formula given in \cite{2015A&A...576A.126A}:

\begin{equation}
\sigma _{F_{var}}=\sqrt{F{_{var}}^{2}+\sqrt{\frac{2}{N}\frac{\left \langle e{_{err}}^{2} \right \rangle^{2}}{\left \langle F \right \rangle^{4}}+\frac{4}{N}\frac{\left \langle \sigma {_{err}}^{2} \right \rangle}{\left \langle F \right \rangle^{2}}F{_{var}}^{2}}}-F_{var}
\end{equation}

In this case, we can get $VA=0.62$, $F_{var}=0.31\pm 0.03$ through the above formula, which shows that there is a modest change in this source during this time. Figure 1 shows the light curve of PKS J0805-0111, and the blue curve is a sinusoidal fitting curve, which is plotted to help to indicate the modulation in the light curve, as can be seen from Figure 1, there is a clear outbreak activity in this source, and four distinct peaks can be clearly seen, from the light curve, we can visually see that this source has a QPO between 1100 days and 1300 days.

\begin{figure}
\centering
\includegraphics[width=1.0\textwidth]{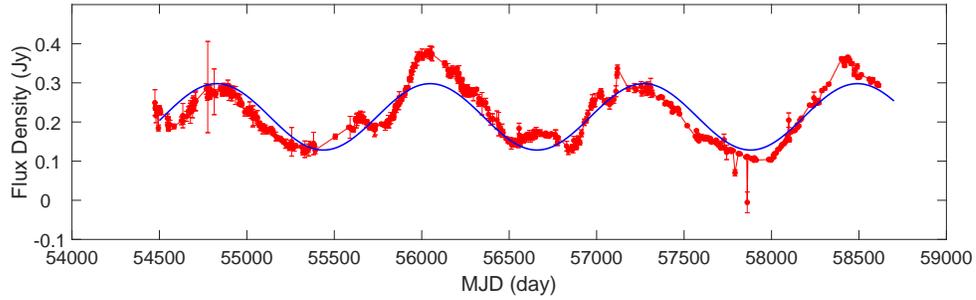}
\vspace{-1em}
\caption{Long-term light curve of FSRQ PKS J0805-0111 at 15 GHz in January 9,2008 to May 9, 2019 (MJD 54474 to 58612) obtained from OVRO, the data integration times is 4138 days ($\sim$11.3years), the blue sinusoidal curve (with a period of 3.38 years) is plotted to help indicate the modulation in the light curve.}
\label{f1}
\end{figure}

\begin{figure}
\centering
\includegraphics[width=0.9\textwidth]{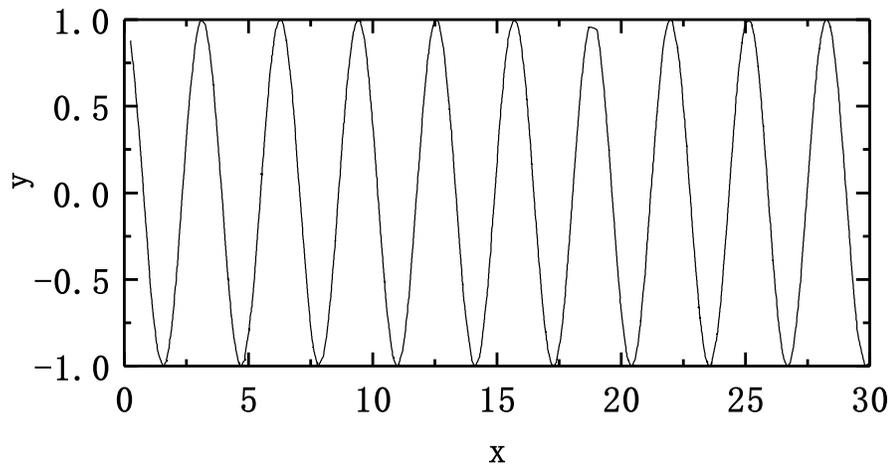}
\vspace{-2em}
\caption{Graph of cosine function analog periodic signals.}
\label{f2}
\end{figure}

\begin{figure}
\centering
\includegraphics[width=0.8\textwidth]{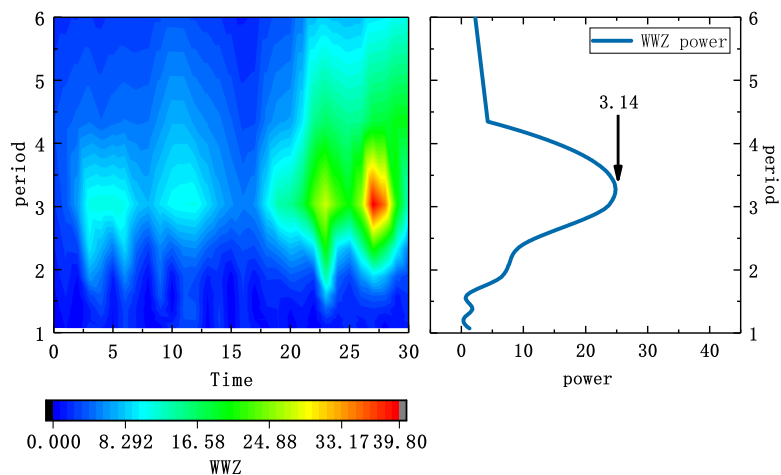}
\caption{Periodic analysis of the cosine function by the WWZ transform.}
\label{f3}
\end{figure}

\begin{figure}
\centering
\includegraphics[width=0.8\textwidth]{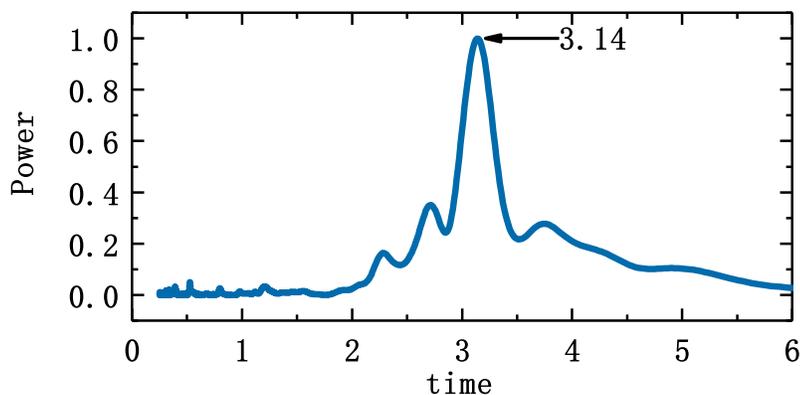}
\caption{ Periodic analysis of the cosine function by the LSP method.}
\label{f4}
\end{figure}

\begin{figure}
\centering
\includegraphics[width=0.8\textwidth]{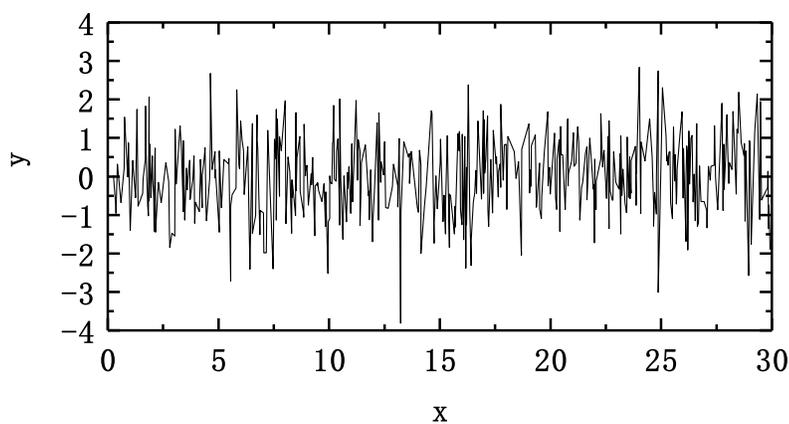}
\caption{random noise.}
\label{f5}
\end{figure}

\begin{figure}
\centering
\includegraphics[width=0.8\textwidth]{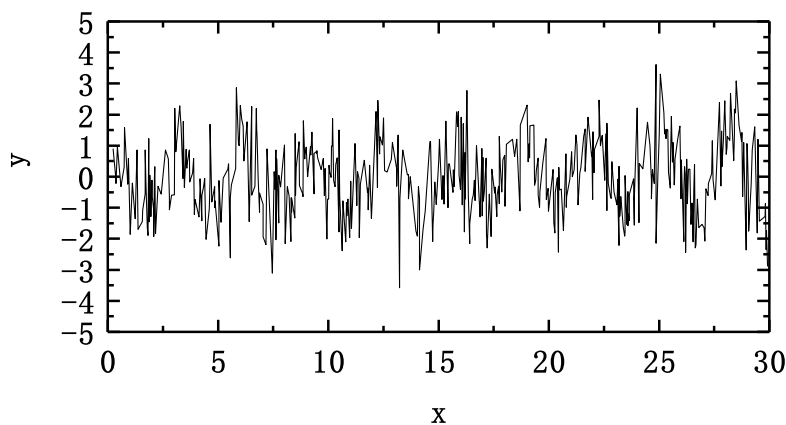}
\caption{Superimposed graph of random noise and cosine function.}
\label{f6}
\end{figure}

\begin{figure}
\centering
\includegraphics[width=0.8\textwidth]{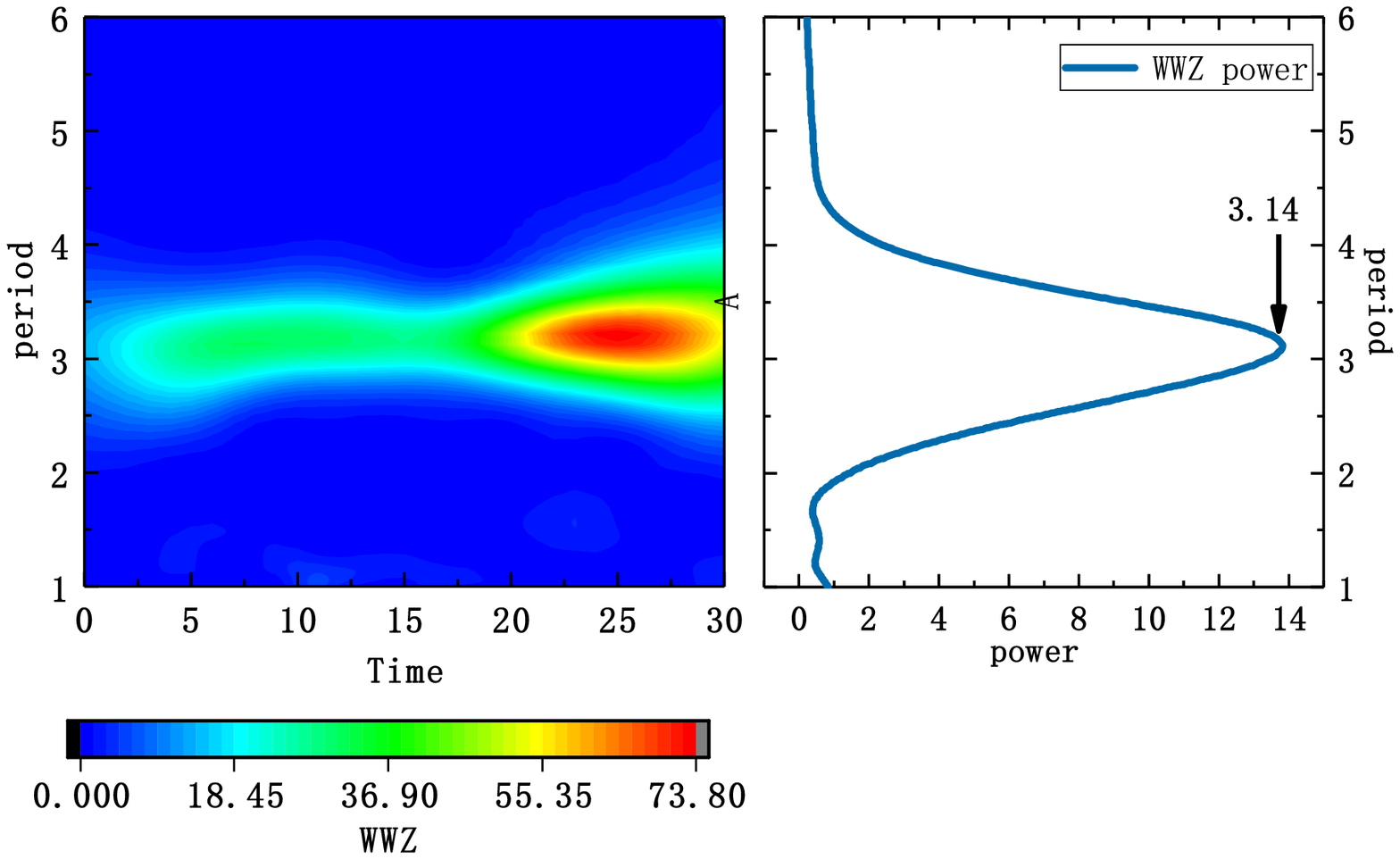}
\caption{Periodic analysis of the random noise and sinusoidal function by the WWZ method.}
\label{f7}
\end{figure}

\begin{figure}
\centering
\includegraphics[width=0.8\textwidth]{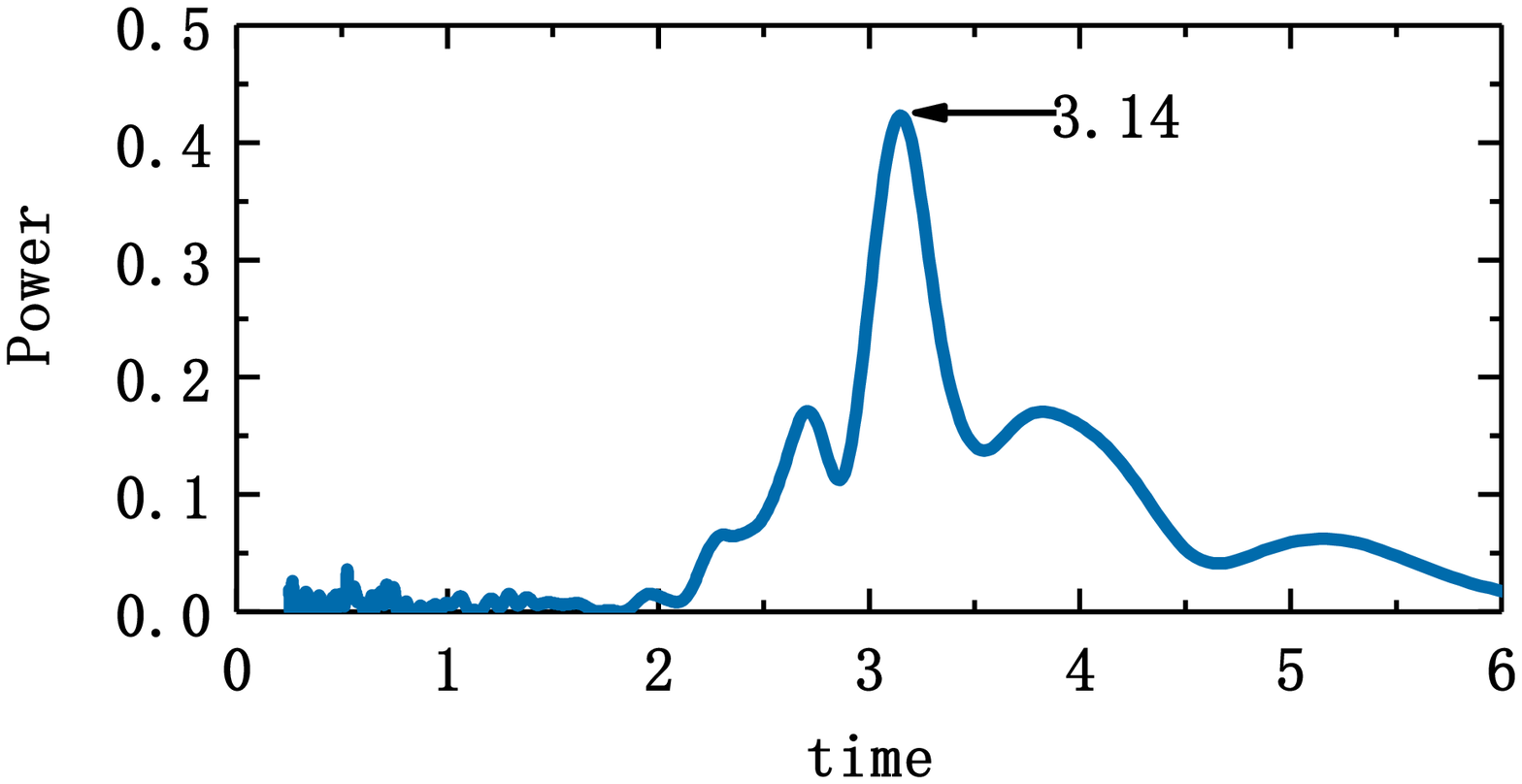}
\caption{Periodic analysis of the random noise and sinusoidal function by the LSP method.}
\label{f8}
\end{figure}

\begin{figure}
\centering
\includegraphics[width=0.8\textwidth]{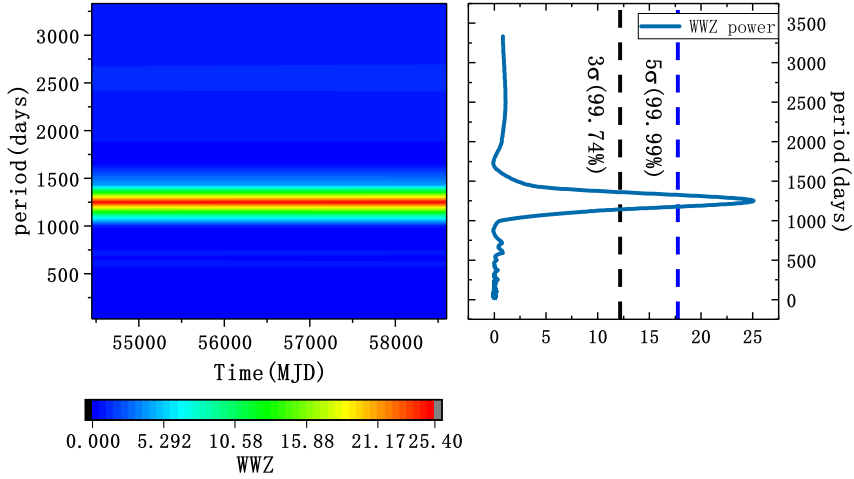}
\caption{WWZ of the light curve presented in Figure 9. The left panel shows the distribution of color-scaled WWZ power (with red most intense and blue lowest) in the time-period plane; the right panel shows the time-averaged WWZ power (solid blue curve) as a function of period showing a distinct peak stands out around the timescale of $1250\pm214$ days; the black dashed lines indicate the thresholds of FAP fixed at $3\sigma \left ( 99.74\% \right )$, the blue dashed lines indicate the thresholds of FAP fixed at $5\sigma \left ( 99.99\% \right )$.}
\label{f9}
\end{figure}

\begin{figure}
\centering
\includegraphics[width=0.8\textwidth]{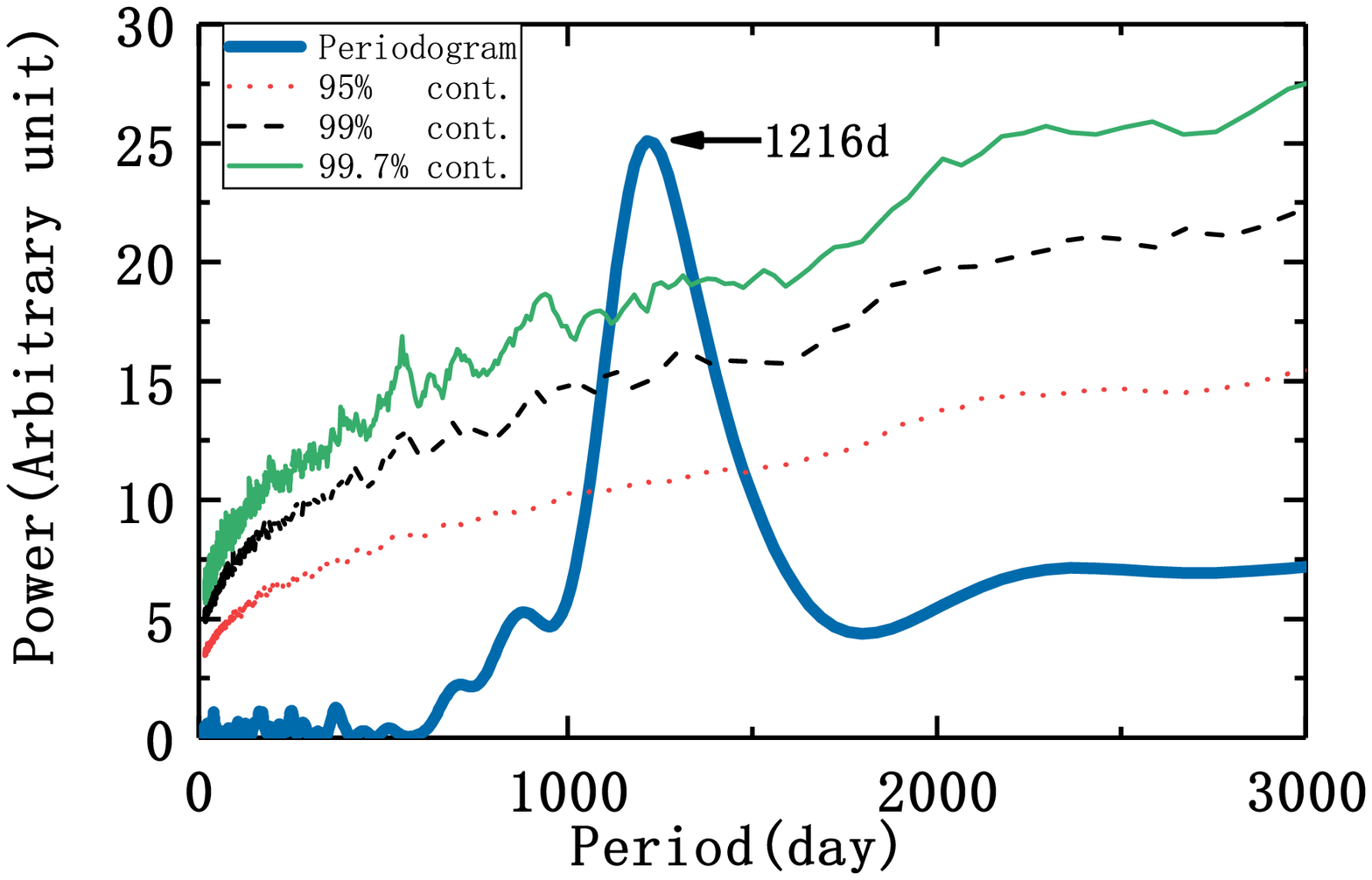}
\caption{The results of the LSP analysis for the period search, showing a distinct peak stands out around the timescale of $1216\pm373$ days. The dashed red, black and green curves represent the $95\%$, $99\%$ and $99.7\%$ significance levels, respectively, from MC simulations.}
\label{f10}
\end{figure}

\section{Light curve analysis and results}
\label{sect:result}

The 15 GHz radio band light curve of FSRQ PKS J0805-0111, for observations taken from January 9,2008 to May 9, 2019 is plotted in Figure. 1. A visual inspection indicated a possible a QPO in the observations made during the whole observations.

We carried out QPO search analysis using two methods of long time-series analysis: WWZ and LSP. We discussed the methods, analysis and the results in detail below.

\subsection{Astronomical analog signal test}

In order to test the reliability of WWZ and LSP method, we used the analog periodic signal as the astronomical observation data for verification. In this paper, we tested the accuracy of the two research methods by the cosine function $y=cos2\pi$ with a period of $\pi$ as shown in Figure 2 , the test results of these two methods are: Figure 3 shows the periodic analysis results of the cosine function by the WWZ method, and the periodic analysis results of the sinusoidal function by the LSP method is shown in Figure 4.

In order to further determine the reliability of the WWZ and LSP methods in analyzing observations, the random noise data are we added to the periodic signal data of the cosine function simulation, and these two methods are tested again. The specific process is calculated by the Python program, and the test results are as follows: Figure 5 shows random noise, the superimposed graph of random noise and cosine function is shown in Figure 6, Figure 7 and Figure 8 shows the results of periodic analysis for the cosine function with random noise added by the WWZ and LSP method, respectively.

The results show that the standard cosine simulation data and the cosine simulation data with random noise was analyzed by WWZ and LSP methods, which is the same as $T\approx 3.14 $, this proves that using these two methods for periodic research is reliable.

\subsection{Weighted Wavelet Z-transform}

Wavelet analysis simultaneously decomposing data into time and frequency domains to estimate and determine the significance of a period \citep{2019MNRAS.484.5785G}. However, we using wavelet analysis to process non-equal interval data in practice, the interpolation method was used to reduce the impact of the astronomical observation signals are affected by the observation season, the weather, and the phase of the moon, but this has a great influence on the authenticity of the data.

We calculated the WWZ power for a given time and frequency by the WWZ method \citep{2013MNRAS.436L.114K, 2016ApJ...832...47B, 2017ApJ...847....7B, 2019MNRAS.484.5785G}. It is defined by \cite{1996AJ....112.1709F}, he points that, the analysis result can be significantly improved, and the period can be obtained more accurately, if the wavelet transform is regarded as the projection of the vector.

\subsection{Lomb-Scargle Periodogram}

Lomb-Scargle Periodogram \citep{1996AJ....112.1709F} is used widely to determine if QPOs are present in the observations, it is a popular method of time series analysis \citep{2016ApJ...832...47B, 2017ApJ...847....7B, 2018Galax...6..136B}. The method is a discrete Fourier transform (DFT)-based periodic extraction algorithm. The LSP's basic principle is that using a series of trigonometric functions by the least-squares method linear combination $y=a cos\omega t+b sin\omega t$ to fit the time series, and the characteristics of signal are converted from the time domain to the frequency domain on this basis \citep{1982ApJ...263..835S}. For the non-uniform sampling time series $x\left ( t_{i} \right )$, $i=1,2,3\cdots ,N$, the power spectrum is defined as:

\begin{equation}
\begin{aligned}
P_{LS}\left ( f \right )=&\frac{1}{2N}\times \left [ \frac{\left \{ \sum_{i=1}^{N}x\left ( t_{i} \right )cos\left [ 2\pi f\left ( t_{i}-\tau  \right ) \right ] \right \}^{2}}{\sum_{i=1}^{N}cos^{2}\left [ 2\pi f\left ( t_{i}-\tau  \right ) \right ]}\right.  \left.+\frac{\left \{ \sum_{i=1}^{N}x\left ( t_{i} \right )sin\left [ 2\pi f\left ( t_{i}-\tau  \right ) \right ] \right \}^{2}}{\sum_{i=1}^{N}sin^{2}\left [ 2\pi f\left ( t_{i}-\tau  \right ) \right ]}\right ]^{2}
\end{aligned}
\end{equation}

The parameter $f$ and $\tau$ represent the test and the time offset, respectively, which can be obtained by the following formula.

\begin{equation}
tan\left ( 2\pi f\tau  \right )=\frac{\sum_{i=1}^{N}sin2\pi ft_{i}}{\sum_{i=1}^{N}cos2\pi ft_{i}}
\end{equation}

\subsection{The variability analysis and results of FSRQ PKS J0805-0111}
In this paper, we analyze the 15 GHz radio observations of the source FSRQ PKS J0805-0111 announced by OVRO used the WWZ and LSP method, we found that the QPO of this source is about 3.38 years.

In order to search for QPOs of the source FSRQ PKS J0805-0111, we analyzed the 15 GHz flux density variations by the WWZ method at first, which is one of the most common time series analysis methods. the WWZ transform of FSRQ PKS J0805-0111 light curve was computed for the minimum and maximum frequencies of $f_{min}=1/4138$, and $f_{max}=1/25$, respectively. The results are shown in Figure 9, The most significant spectral power peak yielded by WWZ, and we further estimate its significance level by testing the false alarm probability (FAP) of the null hypothesis.
The probability that $P_{N}\left ( \omega  \right )=P_{LS}\left ( \omega  \right )/\sigma ^{2}$  will be between some positive $z$ and $z+dz$ is $e^{-z}$.  If we scan some $N_{i}$ independent frequencies, the probability that none give values larger than $z$ is $\left ( 1-e^{-z} \right )^{N_{i}}$: $p\left ( > z \right )\equiv 1-\left ( 1-e^{-z} \right )^{N_{i}}$ is FAP\citep{2004A&A...419..485C}. The smaller the false alarm probability, the higher degree of the significance for the peak.

We analyzed the ¡«11.3 years long OVRO light curve of FSRQ PKS J0805-0111, the results are as follows: It is showing a distinct peak stands out around the timescale of $1250\pm 214$ days with a single-trial FAP significance of $5.75\times 10^{-8}$, which represents a very strong periodicity in the associated periodic signal. We estimated the period uncertainty by calculating the $3\sigma \left ( 99.74\% \right )$ and $5\sigma \left ( 99.99\% \right )$ confidence interval on the observed period, by the FAP method.

We performed the LSP analysis of the entire light curve in order to further confirm the presence of the above QPO by the different method.

We performed LSP method on the observations, the LSP method of FSRQ PKS J0805+0111 light curve was computed for the minimum and maximum frequencies of $f_{min}=1/4138$, and $f_{max}=1/25$, respectively. What needs to be emphasized is that the estimate of the total number of periodogram frequencies $n_{0}$, it is critical to the evaluation of the periodogram. In this work, we evaluate the total number of periodogram frequencies using

\begin{equation}
N_{eval}=n_{0}Tf_{max}
\end{equation}

Since the observations we get are not equally spaced, spurious peaks may be generated, which makes it difficult to estimate the confidence of the peak height in the power spectrum. So, we should consider this effect in this case. The red noise process may be product the periodic variability of blazars, and the red noise was modeled as an approximately power low Power spectral density (PSD): $P\propto f^{-\alpha }+C$. We molding the wavelength variability as red noise with a power law index $\alpha$ to assessed the confidence of our findings. We then performed a Monte Carlo simulation technique \citep{1995A&A...300..707T} for determining the significance of the periods, a large number of (typically 20000) light curves were simulated for every spectral slope $\alpha$ values. We obtaining the $\alpha$ index by fitted the spectrum of the periodogram with power law. Using even sampling interval to simulated the 10000 light curves, and computed their LSP. Figure 10 gives the results, showing a distinct peak stands out around the timescale of $1216\pm 373$ days. The local $95\%$, $99\%$ and $99.7\%$ MC simulations contours are represented by the dashed red, black and the green curves, respectively.

\section{Discussion}
\label{sect:discussion}

Generally, some periodic or quasi-periodic behavior were displayed in the light curves of blazars, studying these periodic or quasi-periodic behavior is an important method to investigate the nature of the physical mechanisms within the emission regions. However, many researchers have been reported the QPOs of a few blazars at different wavelengths on diverse timescales, they can be approximately divided into three classes, viz., intra-day variability, short-term timescale variability, and long-term timescale variability. Studies on QPOs could provide novel insights into a number of blazar aspects, some physical models has been reported by many researchers, e.g., binary SMBH AGN system  \citep{1996ApJ...460..207L,2000ApJ...531..744V,2007A&A...462..547F}, helical structure in inner jets \citep{1993ApJ...411...89C}, precessing accretion disk model \citep{1997ApJ...478..527K}, the thermal instability of thin disks scenario. The observed periodic flux modulations could be explained by a number of models.

In this paper, we have attempted to detect the possible periodicity of the FSRQ PKS J0805-0111 in the radio 15 GHz light curve using the OVRO observations acquired during the period from January 9,2008 to May 9, 2019 (MJD 54474 to 58612). The period of 15 GHz radio band is $1250\pm214$ days by WWZ analysis, the confidence interval on the observed period is more than $5\sigma (99.99\%)$, by the FAP method; the period of 15 GHz radio band by LSP analysis is $1216\pm373$ days, the significance level is more than $99.7\%$, from MC simulations. It can be concluded that the QPO of FSRQ PKS J0805-0111 is $3.38\pm0.8$ years ($> 99.7\%$ confidence level).

Although it is not well understood for the physical mechanism in blazars on long-term timescales, some possible interpretations, such as, 1) the thermal instability of thin disks scenario; 2) the spiral jet scenario; 3) the binary supermassive black hole scenario, have been applied to explain the long-term periodic variability.In the scenario of thermal instability of thin disks, the cyclic periodic outburst leaded by the thermal instability of thin disk. The uncertainty of disk causes the related outburst of the jet because of there is a certain degree of the link between jet and disk. Random light variations could produced by thermal instability of thin disks in jets with stochastic red noise characteristics, this process is stochastic \citep{2017ApJ...847....8L}. Thence, this scenario be responsible for long-term QPO is hard.

In the spiral jet scenario, the relativistic beaming effect lead to the QPO behavior of blazars, the change of relativistic beaming effect cause the arises of the obvious flux variations because of the different parts of such a helical jet pass closest to the line of sight, even though the emission from the jet have no intrinsic variations. Furthermore, the viewing angle to the helical motion changes periodically when the emission blob of the jet moves to us, thereby resulting in QPOs \citep{2018NatCo...9.4599Z}. However, it needs to note that, low-frequency radio emission such as 15 GHz is less affected by the beaming effect, due to it is generally considered to be dominated by the extended jet structure of the jet \citep{2018ApJ...869..133F}.

Consequently, we are more promising that the QPO behavior in the FSRQ PKS J0805-0111 may caused by the binary supermassive black hole model, which possibly can explain presence of year-like long-term timescale variability in AGN \citep{2006MmSAI..77..733K}. A SMBH system would led to a wiggling or precessing jet because of the orbital motion or rotation, i.e., there is a precession motion of a relativistic jet in an orbit due to the gravitational torque induced by the non-coplanar secondary black hole in the primary accretion disc, which produce the observed long-term timescale variabilities ranging from a few to tens of years \citep{1997ApJ...478..527K,2013MNRAS.428..280C}.

If we assume that the long-term light variations of blazars are caused by the binary supermassive black hole model, and the distance between the primary black hole and the secondary black hole can be calculated.For the binary black hole mass ratio, Qian et al. discussed the secondary black hole and primary black hole is about 1 \citep{2007ChJAA...7..364Q}, and Caproni and Abraham think this ratio is about 0.78 \citep{2004ApJ...602..625C}. In this case, we calculated that the average of these two values is 0.89, and we use this value as the ratio of the secondary black hole mass and the primary black hole mass. In addition, we obtain 293 black hole mass of the typical blazars \citep{2011ApJS..194...45S, 2012ApJ...748...49S, 2006ApJ...637..669L, 2004ApJ...615L...9W, 2012ApJ...759..114C, 2012MNRAS.421.1764S, 2009RAA.....9..293Z, 2012ApJ...752..157Z, 1991A&A...249...65X, 2004AJ....127...53X}, and we calculate the average value of these blazars' black hole mass is $\bar{M}=10^{8.6}M_{\bigodot }$. We assume that this is the primary black hole mass, and we take a binary black hole mass ratio of 0.89, therefore, secondary black hole mass can be calculated as $m=0.89\bar{M}=0.89\times 10^{8.6}M_{\bigodot}$. The relevant parameters of the binary black hole model according to the orbit period can calculated.

In this case, we take $P_{obs}=3.38$ years, $M=10^{8.6}M_{\bigodot }$, $m=0.89\times10^{8.6}M_{\bigodot }$, according to Kepler's law:

\begin{equation}
\left ( \frac{P_{obs}}{1+z} \right )^{2}=\frac{4\pi ^{2}a^{3}}{G\left ( M+m \right )},
\end{equation}
where $Z$, $a$ and $G$ represent the red shift, the distance between the primary black hole and the secondary black hole and universal gravitational constant, thus we can calculate the distance between the primary black hole and the secondary black hole is $a\sim 1.71\times 10^{16}$ cm.

\section{Conclusions}
\label{sect:conclusion}

In this paper, we have searched QPOs in the 15 GHz light curve of the FSRQ PKS J0805-0111 monitored by the OVRO 40 m telescope during the period from January 9,2008 to May 9,2019. Our main results are as follows:

(1)We have found a quasi-periodic signal with a period of $3.38\pm0.8$ years ($> 99.7\%$ confidence level) in the 15 GHz radio light curve of the FSRQ PKS J0805-0111 by the WWZ and the LSP method. This is the first time that the quasi-periodic signal has been detected in this source.

(2)In the scenario where the precession of the binary supermassive black holes cause the radio quasi-periodic variability, the distance between the primary black hole and the secondary black hole is about $1.71\times 10^{16}$ cm.

This source could be a good binary supermassive black hole candidate. We will further monitor the optical variability of the source in the optical band to further verify whether there is the precession of binary supermassive black holes.

\begin{acknowledgements}

This work is supported by the National Nature Science Foundation
of China (11663009), and the High-Energy Astrophysics Science and Technology Innovation Team of Yunnan Higher School. This work is supported by the National Nature Science Foundation of China (11663009), and the High-Energy Astrophysics Science and Technology Innovation Team of Yunnan Higher School. This research has made use of data from the OVRO 40-m monitoring program \citep{2018ApJ...869..133F} which is supported in part by NASA grants NNX08AW31G, NNX11A043G, and NNX14AQ89G and NSF grants AST-0808050 and AST-1109911.

\end{acknowledgements}

\label{lastpage}
\bibliographystyle{raa}
\bibliography{bibtex}

\end{document}